\documentclass[12pt,aps,prb]{revtex4}

\usepackage{amsmath}
\usepackage{bm}

\DeclareMathOperator{\csch}{csch}     

\renewcommand{\a}{\alpha}           
\newcommand{\as}{\quad\text{as}\enspace} 
\renewcommand{\b}{\beta}            
\newcommand{\dl}{\delta}            
\newcommand{\dn}{{\mathord{\downarrow}}} 
\newcommand{\E}{\mathcal{E}}        
\newcommand{\half}{\tfrac{1}{2}}    
\newcommand{\ketbra}[2]{|#1\rangle\langle#2|} 
\newcommand{\la}{\lambda}           
\newcommand{\om}{\omega}            
\newcommand{\shalf}{{\scriptstyle\frac{1}{2}}} 
\renewcommand{\th}{\theta}          
\newcommand{\twobytwo}[4]{\begin{pmatrix}#1& #2\\ #3& #4\end{pmatrix}}
\newcommand{\up}{{\mathord{\uparrow}}} 
\newcommand{\vth}{\vartheta}        
\newcommand{\word}[1]{\quad\mbox{#1}\quad} 
\newcommand{\x}{\times}             
\newcommand{\7}{\dagger}            
\renewcommand{\.}{\cdot}            

\newcommand{\gs}{\mathrm{gs}}       
\newcommand{\HF}{\mathrm{HF}}       

\newcommand{\vecform}{\bm}              
\newcommand{\PP}{\vecform{P}}           
\newcommand{\pp}{\vecform{p}}           
\newcommand{\ppp}{\vecform{\pi}}        
\newcommand{\qq}{\vecform{q}}           
\newcommand{\rr}{\vecform{r}}           
\newcommand{\RR}{\vecform{R}}           
\newcommand{\uu}{\vecform{u}}           
\newcommand{\xx}{\vecform{x}}           
\newcommand{\zz}{\vecform{z}}           

\makeatletter
\def\section{\@startsection{section}{1}{\z@}{-3.5ex plus -1ex minus
 -.2ex}{2.3ex plus .2ex}{\large\bfseries}}
\def\subsection{\@startsection{subsection}{2}{\z@}{-3.25ex plus -1ex
 minus -.2ex}{1.5ex plus .2ex}{\normalsize\bfseries}}
\makeatother

\begin{document}

\title{Exact phase space functional for two-body systems}

\author{Jos\'e M. Gracia-Bond\'ia}
\affiliation{Departamento de F\'isica Te\'orica,
Universidad de Zaragoza, 50009 Zaragoza, Spain}
\author{Joseph C. V\'arilly}
\affiliation{Escuela de Matem\'atica, Universidad de Costa Rica,
San Jos\'e 2060, Costa Rica}

\date{21 November 2010}

\begin{abstract}
The determination of the two-body density functional from its one-body
density is achieved for Moshinsky's harmonium model, using a
phase-space formulation, thereby resolving its phase dilemma. The
corresponding sign rules can equivalently be obtained by minimizing
the ground-state energy.
\end{abstract}

\maketitle


\section{Introduction}
\label{sec:intro}

The harmonium model originally proposed by Moshinsky\cite{Moshinsky}
has earned its spurs as a simple analogue to a two-electron atom,
helpful to illustrate the main ideas of reduced density matrix and
correlation energy theory in an exactly solvable context. The model
consists of two spin-$\half$ fermions trapped in a harmonic potential
and repelling each other with a Hooke's law force, as well. Chapter~2
of the book by Davidson\cite{Davidson} describes its ground state in
the standard wave function formalism, as well as the reduced density
matrix and the pair distribution, exhibiting correlation.

A one-dimensional version of the model was put to work by
Neal\cite{Neal} in the hope of finding an exact universal density
functional of the Hohenberg--Kohn--Sham type.\cite{HohenbergK} This
proves illusory; but the computations by Schindlmayr in his very
pedagogical rejoinder\cite{Schindlmayr} make it clear that Neal's
harmonium scheme supports successful approximations for confining
potentials. More recently, the harmonium model has proved its worth in
suggesting approximate general forms of 1-density
matrices\cite{AmoviliM} and Ans\"atze for correlation energy
density.\cite{MarchCCA}

The advantage of the harmonium model is that the required computations can
be analytically performed. However, despite this solvable character,
several pertinent functionals have not been exploited so far. It
is well known that possession of the 1-body matrix~$\rho_1$ for an
$N$-electronic system does not allow effective inference of the
corresponding 2-body matrix~$\rho_2$, which would trivialize the
energy functional in quantum chemistry. It is natural to diagonalize
$\rho_1$ and seek to expand $\rho_2$ in terms of eigenfunctions~$f_j$
of~$\rho_1$ (``natural orbitals'') and its eigenvalues $0\le n_j\le 1$
(``occupation numbers''), with $\sum_jn_j=1$. Over the years, starting
with the work by M\"uller, approximate functionals based on this
spectral analysis of~$\rho_1$ have been suggested and tried with
various results.

Two-electron atoms constitute the exception to our ignorance. In this
article we focus on the \textit{exact} Shull--L\"owdin--Kutzelnigg
(SLK) functional for the ground state of such atoms in terms of
natural orbitals.\cite{Davidson,ColemanY,Piris} Work by
those authors in the late fifties and early sixties established that,
for a reduced 1-density of the kind
$$
\rho_1(\xx,\xx') = \frac{1}{2}\bigl( \up_1\up_{1'} + \dn_1\dn_{1'} \bigr)
\rho_1(\rr,\rr') = \frac{1}{2}\bigl( \up_1\up_{1'} + \dn_1\dn_{1'} \bigr)
\sum_j n_jf_j(\rr)f_j^*(\rr'),
$$
the corresponding 2-density matrix is given by the form
\begin{align}
&\rho_2(\xx_1,\xx_2,\xx'_1,\xx'_2) 
= \frac{1}{2}\bigl( \up_1\dn_2 - \dn_1\up_2 \bigr)
\bigl( \up_{1'}\dn_{2'} - \dn_{1'}\up_{2'})
\label{eq:rho-two} 
\\
&\quad \x \begin{pmatrix}
f_1(\rr_1) & f_2(\rr_1) & f_3(\rr_1) & \cdots
\end{pmatrix} \begin{pmatrix}
c_1 & & & \\ & c_2 & & \\ & & c_3 & \\ & & & \ddots 
\end{pmatrix} \begin{pmatrix}
f_1(\rr_2) \\ f_2(\rr_2) \\ f_3(\rr_2) \\ \vdots \end{pmatrix}
\notag \\
&\quad \x \begin{pmatrix}
f^*_1(\rr'_1) & f^*_2(\rr'_1) & f^*_3(\rr'_1) & \cdots 
\end{pmatrix} \begin{pmatrix}
c_1^* & & & \\ & c_2^* & & \\ & & c_3^* & \\ & & & \ddots
\end{pmatrix} \begin{pmatrix}
f_1^*(\rr'_2) \\ f_2^*(\rr'_1) \\ f_3^*(\rr'_2) \\ \vdots
\end{pmatrix}.
\notag
\end{align}
Alas, the SLK recipe, although exact, is underdetermined: of the $c_j$ we
only know that $|c_j|^2=n_j$. This is a ``phase dilemma'' of density
functional theory. We work here only with states described by real
wavefunctions ---still leaving us with an infinite number of signs to
account~for.

\vspace{6pt}

Notwithstanding its venerable age, formula~\eqref{eq:rho-two}
apparently has never been verified exactly. A theorem without an
example is a sorry thing. Of course, numerical computations tend to
confirm the SLK theorem; but one should not forget that they tell us
about the approximations (nearly always from a Hartree--Fock starting
point), rather than the true solution. We verify the SLK method for
harmonium in full detail, including the energy functional, in the
following three sections. Along the way, we solve the sign conundrum
for the model. Our methods are elementary, asking familiarity with
little more than orthogonal polynomials at the level of
Lebedev\cite{Lebedev} or Andrews \textit{et~al}.\cite{AndrewsAR}

Nevertheless, within the standard formalism it is not at all obvious
how to go about the problem. We manage to sidestep difficulties by
working with the Wigner quasiprobability on phase space instead. A
recent quantum phase space view of harmonium, dealing with other
matters, is given by Dahl.\cite{DahlMosh}

In the concluding Section~5 we very briefly discuss the new
perspectives on correlation energy and approximate functionals
for~$\rho_2$ revealed by the treatment in this paper.

We follow Davidson's notation\cite{Davidson} as much as feasible. A
good review on the M\"uller functional is found in
Ref.~\onlinecite{FrankLSS}. One may consult Refs.\
\onlinecite{Helbig}, \onlinecite{LHGross} for popular variations
on~it.

\section{The setup}
\label{sec:vea}

The Hamiltonian for harmonium in Hartree units is
\begin{equation}
H = \frac{p_1^2}{2} + \frac{p_2^2}{2} + \frac{k}{2}(r_1^2 + r_2^2)
- \frac{\dl}{4} r^2_{12}.
\label{eq:Mosh-atom} 
\end{equation}
Introduce extracule and intracule coordinates, respectively given by
$$
\RR = (\rr_1 + \rr_2)/\sqrt{2}, \qquad \rr = (\rr_1 - \rr_2)/\sqrt{2},
$$
with conjugate momenta
$$
\PP = (\pp_1 + \pp_2)/\sqrt{2}, \qquad \pp = (\pp_1 - \pp_2)/\sqrt{2}.
$$
Therefore
$$
H = H_R + H_r = \frac{P^2}{2} + \frac{\om^2 R^2}{2} + \frac{p^2}{2}
+ \frac{\mu^2r^2}{2}.
$$
As advertised, our notation is that of Davidson~\cite{Davidson} except
that our $\dl$ is equal to twice his $\a$, and we introduce the
frequencies $\om = \sqrt k$ and $\mu = \sqrt{k-\dl}$; assume
$0 \leq \dl < k$ for both particles to remain in the potential well.

Since the spin factors are known, we concentrate on the spinless part
of the quantum states henceforth. The spinless Wigner quasiprobability
(normalized to one) corresponding to a (real) 2-particle wave function
$\Psi$ is given by
\begin{equation}
P_\Psi(\rr_1,\rr_2;\pp_1,\pp_2) = \frac{1}{\pi^6} \int
\rho_2(\rr_1 - \zz_1, \rr_2 - \zz_2; \rr_1 + \zz_1, \rr_2 + \zz_2)\,
e^{2i(\pp_1\.\zz_1 + \pp_2\.\zz_2)} \,d^3z_1 \,d^3z_2,
\label{eq:phase-two} 
\end{equation}
with $\rho_2(\rr_1,\rr_2,\rr'_1,\rr'_2) =
\Psi(\rr_1,\rr_2) \Psi(\rr'_1,\rr'_2)$. The definition extends to
transition matrices $\ketbra{\Phi}{\Phi'}$ also:
$$
P_{\Phi\Phi'}(\rr_1,\rr_2;\pp_1,\pp_2) = \frac{1}{\pi^6} \int
\Phi(\rr_1 - \zz_1, \rr_2 - \zz_2) \Phi'(\rr_1 + \zz_1, \rr_2 + \zz_2)
\,e^{2i(\pp_1\.\zz_1 + \pp_2\.\zz_2)} \,d^3z_1 \,d^3z_2.
$$
Fourier analysis easily provides the inverse formula to this, that we
do not bother to write.

By use of~\eqref{eq:phase-two} and the ground state wave function for
harmonium, one can obtain the corresponding Wigner function, which
factorizes into extracule and intracule parts:
\begin{equation}
P_\gs(\rr_1,\rr_2;\pp_1,\pp_2) = \frac{1}{\pi^6}
\exp\biggl(-\frac{2H_R}\om \biggr) \exp\biggl( -\frac{2H_r}\mu
\biggr).
\label{eq:west-ham} 
\end{equation}
This is reached more efficiently and elegantly by the methods of
phase space quantum mechanics.\cite{Callisto} One can now obtain
$\rho_2$, given by the inverse formula of~\eqref{eq:phase-two}.
The pairs density $\rho_2(\rr_1,\rr_2,\rr_1,\rr_2)$ is recovered by
integration over the momenta.

The reduced 1-body phase space (spinless) quasidensity for the ground
state $d_\gs$ is obtained, as in the standard formalism, by
integrating out one set of variables,
\begin{equation}
d_\gs(\rr;\pp) = \frac{2}{\pi^3}
\biggl( \frac{4\om\mu}{(\om + \mu)^2} \biggr)^{3/2}
e^{-2r^2\om\mu/(\om + \mu)} e^{-2p^2/(\om + \mu)}.
\label{eq:fair-ground} 
\end{equation}
We leave it as an exercise for the reader to recover Eq.~(2--68) of
Ref.~\onlinecite{Davidson} for $\rho_1(\rr_1,\rr'_1)$ from this. The
marginals of $d_\gs$ give the electronic density and momentum density.

It should be recognized that, while $P_\gs$ is a pure state,
mathematically $d_\gs$ describes a mixed state. For Gaussians on phase
space,  such as $P_\gs$ and $d_\gs$~too, there are simple rules to
determine whether they represent a pure state,\cite{Robert} a mixed
state,\cite{Titania} or neither. Writing
$\qq = (\rr_1,\rr_2)$, $\ppp = (\pp_1,\pp_2)$, $\uu = (\qq,\ppp)$,
we find $P_\gs(\uu) = \pi^{-6} e^{-\uu\.F\uu} =
\pi^{-6} e^{-\qq\.A\qq - \ppp\.A^{-1}\ppp}$ where, amusingly,
\begin{equation}
A = \frac{1}{2} \twobytwo{\om + \mu}{\om - \mu}{\om - \mu}{\om + \mu},
\qquad 
A^{-1} = \frac{1}{2} \twobytwo{\om^{-1} + \mu^{-1}}{\om^{-1} - \mu^{-1}}
{\om^{-1} - \mu^{-1}}{\om^{-1} + \mu^{-1}}.
\label{eq:topsy-turvy} 
\end{equation}
We see that the matrix $F$ corresponding to formula~\eqref{eq:west-ham}
is symmetric and symplectic, and therefore represents a pure state.
This is not the case for~$d_\gs$. Thus recovering $P_\gs$ from
knowledge of~$d_\gs$ alone is akin to putting Humpty Dumpty together
again!

\section{Computation of the 2-body quasidensity}
\label{sec:oiga}

Since all the relevant quantities factorize, in this section we work
in one dimension (instead of three) for notational simplicity. The
real quadratic form in the exponent of $d_\gs$ must be symplectically
congruent to a diagonal one.\cite{Titania} We perform the
transformation
$$
(Q,P) := \bigl((\om\mu)^{1/4}r, (\om\mu)^{-1/4}p\bigr);
\word{or, in shorthand,}  U = Su,
$$
where now $u = (r,p)$. Here $S$ being symplectic just means having
determinant~$1$, which is evidently the case. Introducing as well the
parameter $\la := 2\sqrt{\om\mu}/(\om + \mu)$, the $1$-quasidensity
takes the simple form
$$
d_\gs(u(U)) = \frac{\la}{\pi}\, e^{-\la U^2}.
$$
We may also write $\la =: \tanh(\b/2)$, so that
\begin{gather*}
\b = \log \frac{1 + \la}{1 - \la}
= 2\log \frac{\sqrt\om + \sqrt\mu}{\sqrt\om - \sqrt\mu},  \word{and}
\sinh(\b/2) = \frac\la{\sqrt{1 - \la^2}}
= \frac{2\sqrt{\om\mu}}{\om - \mu}.
\end{gather*}

{}From the series formula, valid for $|t| < 1$,
$$
\sum_{n=0}^\infty L_n(x)\, e^{-x/2}\, t^n
= \frac{1}{(1 - t)}\, e^{-x(1+t)/2(1-t)},
$$
taking $t = -(1 - \la)/(1 + \la) = - e^{-\b}$ and $x = 2U^2$, it
follows that
$$
\frac{\la}{\pi}\, e^{-\la U^2} = \frac{2}{\pi}\, \sinh \frac{\b}{2}
\sum_{r=0}^\infty (-1)^r L_r(2U^2)\, e^{-U^2} e^{-(2r+1)\b/2}.
$$
We recognize the basis of Wigner eigenfunctions on phase space
standing for the oscillator states:\cite{Callisto}
$$
f_{rr}(U) = \frac{1}{\pi}\,(-1)^r L_r(2U^2)\, e^{-U^2}.
$$
Note the normalization $\int f^2_{rr}(Q,P)\,dU = (2\pi)^{-1}$.
Consequently, we realize that $d_\gs$ is in thin disguise a
\textit{Gibbs state},\cite{Titania} with inverse temperature~$\b$:
\begin{equation}
d_\gs(u) = d_\gs(S^{-1}U)
= 2\sinh \frac{\b}{2} \sum_{r=0}^\infty e^{-(2r+1)\b/2} f_{rr}(U).
\label{eq:high-ground} 
\end{equation}
Thus we have identified the natural orbitals in the $U$~variables.
Their occupation numbers are
\begin{equation}
n_r = 2\sinh \frac\b2\, e^{-(2r+1)\b/2}
= \frac{4\sqrt{\om\mu}}{\om - \mu}\, \biggl(
\frac{\sqrt\om - \sqrt\mu}{\sqrt\om + \sqrt\mu} \biggr)^{2r+1}
= \frac{4\sqrt{\om\mu}}{(\sqrt\om + \sqrt\mu\,)^2}\,
\biggl( \frac{\sqrt\om - \sqrt\mu}{\sqrt\om + \sqrt\mu} \biggr)^{2r}.
\label{eq:power-tool} 
\end{equation}
Notice that $n_0 = 1 - e^{-\b} = Z^{-1}(\b)$, where $Z$ is the
partition function for the system; also
$\sum_r n_r = (1 - e^{-\b}) \sum_r e^{-r\b} = 1$. These $n_r$ have nice
square roots:
$$
\sqrt{n_r} = \frac{2(\om\mu)^{1/4}}{\sqrt\om + \sqrt\mu}\,
\biggl( \frac{\sqrt\om - \sqrt\mu}{\sqrt\om + \sqrt\mu} \biggr)^r.
$$

We prepare now to test the SLK functional. On phase space,
formula~\eqref{eq:rho-two} is replaced by
\begin{equation}
P_{2\,\mathrm{SLK}}(u_1,u_2;\mathrm{spin})
= \frac{1}{2}\bigl( \up_1\dn_2 - \dn_1\up_2 \bigr)
\bigl( \up_{1'}\dn_{2'} - \dn_{1'}\up_{2'})
\sum_{r,s=0}^\infty c_rc_s\, f_{rs}(u_1) f_{rs}(u_2).
\label{eq:silk-road} 
\end{equation}
The $f_{rs}$ are Wigner eigentransitions, the functions on phase space
corresponding to matrix transitions between oscillator states. They
are well known.\cite{Callisto} For $r\ge s$, abusing notation,
$$
f_{rs}(u) := \frac{1}{\pi}\,(-1)^s \sqrt{\frac{s!}{r!}} \,
(2U^2)^{(r-s)/2} e^{-i(r-s)\vth} L_s^{r-s}(2U^2) \, e^{-U^2},
$$
where $\vth := \arctan(P/Q)$. Then $f_{sr}$ is the complex conjugate
of~$f_{rs}$. In~\eqref{eq:silk-road} we proceed to sum over each
subdiagonal, where $r - s = l \geq 0$:
\begin{align*}
\sum_{r-s=l} & \sqrt{n_r n_s} f_{rs}(u_1) f_{rs}(u_2) 
\\
&= \frac{n_0}{\pi^2}\, e^{-l\b/2} (2U_1U_2)^l e^{-il(\vth_1 + \vth_2)}
\,e^{-U_1^2-U_2^2} \sum_{s=0}^\infty \frac{s!}{(l + s)!}\,
e^{-s\b}\, L_s^l(2U_1^2) L_s^l(2U_2^2)
\\
&= \frac{1}{\pi^2}\, e^{-(U_1^2 + U_2^2)/\la}\,
e^{-il(\vth_1 + \vth_2)} I_l\biggl( \frac{2U_1U_2}{\sinh(\b/2)}
\biggr),
\end{align*}
where $I_l$ denotes the modified Bessel function, on use of another 
series formula:\cite{Lebedev}
$$
\sum_{n=0}^\infty \frac{n!}{(n + \a)!} L_n^\a(x) L_n^\a(y)\, t^n
= \frac{(xyt)^{-\a/2}}{1 - t}\, e^{-(x+y)t/(1-t)}\,
I_\a\biggl( \frac{2\sqrt{xyt}}{1 - t} \biggr).
$$
Similarly for $r - s = -l < 0$, we get the same result replaced by
its complex conjugate. Borrowing finally the generating function
identity for Bessel functions,
$$
I_0(z) + 2 \sum_{l=1}^\infty I_l(z) \cos(l\th) = e^{z\cos\th},
$$
where, by taking $\th = \vth_1 + \vth_2 + \pi$, one obtains for the
total sum:
\begin{align*}
& \pi^{-2}
e^{-[(U_1^2 + U_2^2)/\la + 2U_1U_2\csch(\b/2)\cos(\vth_1 + \vth_2)]}
\\
&\quad = \pi^{-2}
e^{-\shalf[(q_1^2 + q_2^2)(\om + \mu)
+ (p_1^2 + p_2^2)(\om^{-1} + \mu^{-1})]} e^{-q_1q_2(\om - \mu)}
e^{p_1p_2(\mu^{-1} - \om^{-1})},
\end{align*}
which in view of~\eqref{eq:west-ham} is the correct result. Clearly
the choice $\th = \vth_1 + \vth_2 + \pi$ amounts to the
\textit{alternating sign} rule for the functional:
$$
c_r = (-1)^r \,\sqrt{n_r}  \word{for}  r = 0,1,2,\dots
$$
In the end, for $P_{2\,\mathrm{SLK}}(u_1,u_2;\mathrm{spin})$ we obtain:
$$
\frac{1}{2}\bigl( \up_1\dn_2 - \dn_1\up_2 \bigr)
\bigl( \up_{1'}\dn_{2'} - \dn_{1'}\up_{2'}) \sum_{r,s=0}^\infty
(-)^{n_r+n_s} \sqrt{n_rn_s}\, f_{rs}(u_1) f_{rs}(u_2).
$$

As far as we know, this is the first time that the solution to the
sign dilemma has been exhibited for any model. No big deal, a devil's
advocate might say, since $P_\gs$ was known beforehand. But, in point
of fact, the correct choice of signs may instead be chosen by
optimization of the \textit{energy functional}; so it can be found
without prior knowledge of the system's ground state. Our next step is
to confirm this claim.

\section{The energy functional}
\label{sec:entienda}

We still work in dimension one. The energy $E_\gs$ of the ground state
is of course $\om/2+\mu/2$. This contains one-body contributions
$E_{1\gs}$ and two-body contributions $E_{2\gs}$. The former
correspond to the kinetic and confinement energy parts. Remember first
that the $1$-body Hamiltonian is given by
$$
h(u) = \frac{p^2}{2} + \frac{\om^2 r^2}{2}
= \sqrt{\om\mu} \biggl( \frac{P^2}{2} + \frac{\om Q^2}{2\mu} \biggr).
$$
It is a simple exercise to obtain $E_{1\gs}$ by integration of
expression~\eqref{eq:fair-ground} with this observable:
$$
E_{1\gs} = \frac{\om}{2} + \frac{\mu + \om^2/\mu}{4}.
$$
More instructive is to prove that this equals $2 \sum_r n_rh_{rr}$,
where $h_{rr}$ denotes the $1$-body energy associated to each natural
orbital. The calculation runs as follows:
\begin{align*}
2\sum_r n_rh_{rr} 
&= \frac{8\,\om\mu}{(\sqrt\om + \sqrt\mu\,)^2}\, \sum_{r=0}^\infty
\biggl( \frac{\sqrt\om - \sqrt\mu}{\sqrt\om + \sqrt\mu} \biggr)^{2r}
\int f_{rr}(Q;P) \biggl( \frac{P^2}{2} + \frac{\om Q^2}{2\mu} \biggr)
\,dQ\,dP
\\
&= \frac{2\,\om\mu}{(\sqrt\om + \sqrt\mu\,)^2}\,
\biggl( 1 + \frac{\om}{\mu} \biggr) \sum_{r=0}^\infty (2r + 1)
\biggl( \frac{\sqrt\om - \sqrt\mu}{\sqrt\om + \sqrt\mu} \biggr)^{2r}
\\
&= \frac{2\,\om\mu}{(\sqrt\om + \sqrt\mu\,)^2}\,
\biggl( 1 + \frac{\om}{\mu} \biggr)
\frac{2(\om + \mu)(\sqrt\om + \sqrt\mu\,)^2}{16\om\mu}
= \frac{\om}{2} + \frac{\mu + \om^2/\mu}{4}.
\end{align*}
We have used~\eqref{eq:power-tool} and the identities
$$
\int f_{rr}(Q;P) P^2 \,dQ\,dP = \int f_{rr}(Q;P) Q^2 \,dQ\,dP
= r + \frac{1}{2} \,; \qquad
\sum_{r=0}^\infty (2r + 1) x^r = \frac{1 + x}{(1 - x)^2} \,.
$$

Now for the two-body contributions. The interelectronic repulsion
potential in~\eqref{eq:Mosh-atom} is
$\dfrac{\mu^2-\om^2}{4}\,r_{12}^2$, so these contributions are of the
form $\sum_{rs} c_r c_s L_{sr}$, with the $L_{sr}$ given by:
\begin{align*}
L_{sr} &= \frac{\mu^2 - \om^2}{4} \int f_{sr}(q_1;p_1) f_{sr}(q_2;p_2)
(q_1 - q_2)^2 \,dq_1\,dq_2 \,dp_1\,dp_2
\\
&= \frac{\mu^2 - \om^2}{4\sqrt{\om\mu}} \int h_s(Q_1) h_r(Q_1)
(Q_1 - Q_2)^2 h_s(Q_2) h_r(Q_2) \,dQ_1\,dQ_2.
\end{align*}
Here $h_r$ are the usual harmonic oscillator eigenfunctions for unit
frequency. We consider the diagonal $r = s$ first, whereby
\begin{align*}
L_{rr} &= \frac{\mu^2 - \om^2}{2\sqrt{\om\mu}} 
\biggl( r + \frac{1}{2} \biggr);  \word{and thus}
\\[\jot]
\sum_r n_r L_{rr} 
&= \frac{\mu^2 - \om^2}{2\sqrt{\om\mu}}\,
\frac{4\sqrt{\om\mu}}{(\sqrt\om + \sqrt\mu\,)^2}\,
\frac{(\om + \mu)(\sqrt\om + \sqrt\mu\,)^2}{16\om\mu}
= \frac{\mu^2 - \om^2}{4\mu}\, \frac{\om + \mu}{2\om} \,.
\end{align*}
We have used that the expected value of $Q^2$ for a harmonic
oscillator eigenstate is $r + \half$ and that the expected value
of~$Q$ is zero. Notice that $\dfrac{\om + \mu}{2\om} < 1$. Therefore,
to fill up the presumed energy gap $(\om^2 - \mu^2)/4\mu$ we have to
``dig deeper''.

\vspace{6pt}

Now $\int h_s(Q) h_r(Q) \,dQ = 0$ for $s \neq r$, by orthogonality. A
non-vanishing contribution of the off-diagonal part may then come 
(only) from the terms
$$
\pm \frac{\om^2 - \mu^2}{2\sqrt{\om\mu}} \sqrt{n_r n_{r+1}}
\biggl[ \int h_r(Q) h_{r+1}(Q) Q \,dQ \biggr]^2.
$$
We compute:
\begin{align*}
\pm \frac{\om^2 - \mu^2}{\sqrt{\om\mu}} 
& \sum_{r=0}^\infty \sqrt{n_r n_{r+1}}
\biggl[ \int h_r(Q) h_{r+1}(Q) Q \,dQ \biggr]^2
\\
&= \pm \frac{\om^2 - \mu^2}{\sqrt{\om\mu}}\,
\frac{4\sqrt{\om\mu}(\sqrt\om - \sqrt\mu\,)}{(\sqrt\om + \sqrt\mu\,)^3}
\sum_{r=0}^\infty \biggl(
\frac{\sqrt\om - \sqrt\mu}{\sqrt\om + \sqrt\mu} \biggr)^{2r}
\frac{r + 1}{2}
\\
&= \pm(\om^2 - \mu^2)\,
\frac{2(\sqrt\om - \sqrt\mu\,)}{(\sqrt\om + \sqrt\mu\,)^3}\,
\frac{(\sqrt\om + \sqrt\mu\,)^4}{16\,\om\mu}
\\
&= \pm \frac{\om^2 - \mu^2}{4\mu}\, \frac{\om - \mu}{2\om} \,.
\end{align*}
Here we employ $\sum_{r=0}^\infty (r + 1)x^r = (1 - x)^{-2}$. The
factor $(r + 1)/2$ comes from the definition of the emission 
operators $a^\7 = (Q - iP)/\sqrt2$ (or the absorption operators), with
$a^\7 h_r = \sqrt{r + 1}\,h_{r+1}$. There is also an overall factor
of~$2$ coming from two subdiagonals for each~$r$.

Obviously there are no other contributions. In order to minimize the
energy we now have to choose \textit{minus} signs whenever
$s = r \pm 1$, so our contention on the alternating sign rule in the
SLK functional for the harmonium model is proved; indeed, in this
section we made no use of $P_\gs$ whatsoever. The total energy comes
out as
$$
\frac{\om}{2} + \frac{\mu + \om^2/\mu}{4}
+ \frac{\mu^2 - \om^2}{4\mu}\, \frac{\om + \mu}{2\om}
- \frac{\om^2 - \mu^2}{4\mu}\, \frac{\om - \mu}{2\om}
= \frac{\om}{2} + \frac{\mu}{2},
$$
as it ought to be.

\section{Discussion}
\label{sec:piensa}

That's all very well, the devil's advocate now concedes. But is it not
rather baroque? At the heart of density functional theory there is the
proof of existence of a functional yielding the ground state energy
from $d_\gs$. We have managed to get it by a roundabout method
equivalent to reconstructing the two-body state. Can't we just proceed
directly? Yes, we can: the energy of the ground state is just (twice)
the average energy of the Gibbs ensemble\cite{BartlettM} represented
by~\eqref{eq:high-ground}. To~wit,
$$
E_\gs = E[d_\gs]
= 2\sqrt{\om\mu} \biggl( \frac{1}{e^\b - 1} + \frac{1}{2} \biggr)
= \frac{\om + \mu}{2}.
$$

\vspace{6pt}

Nearly all the exchange-correlation energy functionals currently used
in quantum chemistry trace their ancestry to that of
M\"uller.\cite{FrankLSS,Piris,Helbig,LHGross,GritsenkoPB} Such
approximations, written in our terms, are most often of the following
form:
$$
\E_\mathrm{xc}[d] = - \frac{1}{2} \sum_{j,k=1}^\infty a(n_j,n_k)
\int f_{jk}(\xx_1) V\bigl(|\rr_1 - \rr_2|\bigr) f_{kj}(\xx_2)
\,d\xx_1 \,d\xx_2,
$$
with integration both on spin and orbital variables. These are all
actually recipes for~$d_2$. For the M\"uller functional
$a(n_j,n_k) = \sqrt{n_jn_k}$. A handy list of alternatives is provided
in Ref.~\onlinecite{Helbig}. According to that reference, all of them
(except for the Hartree--Fock functional) violate antisymmetry; nearly
all of them violate the sum rule for~$d_2$; as well as invariance
under exchange of particles and holes for the correlation part.

It is well known that the M\"uller functional is
overbinding. Our own rigorous proof of this fact for real two-electron
atoms\cite{Pluto} is much more transparent than the one in
Ref.~\onlinecite{FrankLSS} and shows that definite positivity of the
Coulomb potential does play a decisive role, whereas the extra minus
signs in M\"uller's functional do not. For these very reasons the
M\"uller functional's tendency to overcorrelate needs reexamination in
harmonium. Differences between Coulomb and confining potentials are of
course considerable; nevertheless, detailed analytic comparison of the
proposed functionals with the exact one remains an useful exercise,
throwing some light, from our viewpoint, on the elusive correlation
functional. This will be done elsewhere.

Also, the remark at the beginning of this section pictures the
harmonium ``atom'' as a system in equilibrium, with temperature
depending on the strength of the interelectronic repulsion. Although
matters are very different for confining potentials versus
electrostatic ones, as well as for atoms with more than two electrons,
it would seem to suggest that concentration on $\E_\mathrm{xc}[d]$ is
a poor strategy.

\vspace{6pt}

The Wigner function for the Hartree--Fock state for harmonium is given
by the quasiprobability
\begin{align*}
P_\HF(\rr_1,\rr_2;\pp_1,\pp_2)
&= \frac{1}{\pi^6}\, e^{-(r_1^2 + r_2^2)\sqrt{(\om^2 + \mu^2)/2}}
\, e^{-(p_1^2 + p_2^2)/\sqrt{(\om^2 + \mu^2)/2}}
\\
&= \frac{1}{\pi^6}\, e^{-(R^2 + r^2)\sqrt{(\om^2 + \mu^2)/2}}
e^{-(P^2 + p^2)/\sqrt{(\om^2 + \mu^2)/2}},
\end{align*}
so that the expressions~\eqref{eq:topsy-turvy} are replaced by the 
rather trivial
$$
A = \twobytwo{\sqrt{(\om^2 + \mu^2)/2}}{}{}{\sqrt{(\om^2 + \mu^2)/2}},
\quad 
A^{-1} = \twobytwo{1/\sqrt{(\om^2 + \mu^2)/2}}{}{}
{1/\sqrt{(\om^2 + \mu^2)/2}}.
$$
Coming to the correlation energy for harmonium: use of $P_\HF$ gives
\begin{align*}
E_\HF &= \frac{1}{\pi^6} \int \biggl( \frac{P^2}{2}
+ \frac{\om^2 R^2}{2} + \frac{p^2}{2} + \frac{\mu^2 r^2}{2} \biggr)
\\
&\qquad \x e^{-(R^2 + r^2)\sqrt{(\om^2 + \mu^2)/2}}\,
e^{-(P^2 + p^2)/\sqrt{(\om^2 + \mu^2)/2}}
\,d^3\PP \,d^3\RR \,d^3\pp \,d^3\rr
\\
&= \frac{3\sqrt{(\om^2 + \mu^2)/2}}{4}
+ \frac{3\,\om^2}{4\sqrt{(\om^2 + \mu^2)/2}}
+ \frac{3\sqrt{(\om^2 + \mu^2)/2}}{4}
+ \frac{3\mu^2}{4\sqrt{(\om^2 + \mu^2)/2}}
\\
&= 3\sqrt{(\om^2 + \mu^2)/2},
\end{align*}
and so the correlation energy is
\begin{align*}
E_c := E_0 - E_\HF = 3 \biggl( \frac{\om + \mu}{2}
- \sqrt{\frac{\om^2 + \mu^2}{2}} \,\biggr)
\sim - \frac{3\,\dl^2}{32\,\om^3} \as \dl \downarrow 0.
\end{align*}
March and coworkers have suggested a definition for the correlation
energy density on configuration space on the basis of the
Hartree--Fock wave function and the exact ground state for harmonium.
However, relative momentum is as important as relative position in
determining interelectronic correlation, and it seems more appealing
to define a correlation energy density on \textit{phase space}, in the
spirit of Rassolov\cite{Rassolov} and of more recent work by Gill and
coworkers.\cite{GillCOB} We deal with this elsewhere.


\begin{thebibliography}{23}
	
\bibitem{Moshinsky}
M. Moshinsky,
``How good is the Hartree--Fock approximation'',
Am. J. Phys. \textbf{36}, 52--53 (1968).

\bibitem{Davidson}
E. R. Davidson,
\textit{Reduced Density Matrices in Quantum Chemistry},
Academic Press, New York,~1976.

\bibitem{Neal}
H. L. Neal,
``Density functional theory of one-dimensional two-particle systems'',
Am.~J. Phys. \textbf{66}, 512--516 (1998).

\bibitem{HohenbergK}   
P. Hohenberg and W. Kohn,
``Inhomogeneous electron gas'',
Phys. Rev. \textbf{136}, B864--871 (1964); 
W. Kohn and L. Sham,
``Self-consistent equations including exchange and correlation 
effects'', \textit{ibidem} \textbf{140}, A1133--1138 (1965).

\bibitem{Schindlmayr}
A. Schindlmayr,
``Universality of the Hohenberg--Kohn functional'',
Am. J. Phys. \textbf{67}, 933--934 (1999).

\bibitem{AmoviliM}
C. Amovili and N. H. March,
``Exact density matrix for two-electron model atom and approximate 
proposals for realistic two-electron systems'',
Phys. Rev. A \textbf{67}, 022509 (2003).

\bibitem{MarchCCA}
N. H. March, A. Cabo, F. Claro and G. G. N. Angilella,
``Proposed definitions of the correlation energy from a Hartree--Fock 
starting point: the two-electron Moshinsky model as an exactly 
solvable model'',
Phys. Rev. A \textbf{77}, 042504 (2008).

\bibitem{ColemanY}
A. J. Coleman and V. I. Yukalov,
\textit{Reduced Density Matrices. Coulson's Challenge},
Springer, Berlin, 2000.

\bibitem{Piris}
M. Piris,
``Natural orbital functional theory'',
in \textit{Reduced-Density-Matrix Mechanics},
Adv. Chem. Phys. \textbf{134}, D. A. Mazziotti and S. A. Rice, eds.,
Wiley, Hoboken, NJ, 2007; pp.~387--427.

\bibitem{Lebedev}
N. N. Lebedev,
\textit{Special Functions and their Applications},
Dover, New York, 1972.

\bibitem{AndrewsAR}
G. E. Andrews, R. Askey and R. Roy,
\textit{Special Functions},
Cambridge University Press, Cambridge, 1999.

\bibitem{DahlMosh}
J. P. Dahl,
``Moshinsky atom and density functional theory. A phase space view'',
Can.~J. Chem. \textbf{87}, 784--789 (2009).

\bibitem{FrankLSS}
R. L. Frank, E. H. Lieb, R. Seiringer and H. Siedentop,
``M\"uller exchange-correlation energy in density-matrix functional theory'',
Phys. Rev. A \textbf{76}, 052517 (2007).

\bibitem{Helbig}
N. Helbig,
``Orbital functionals in density-matrix and current-density 
functional theory'',
Doktorarbeit, Freie Universit\"at Berlin, 2006.

\bibitem{LHGross}
N. N. Lathiotakis, N. Helbig and E. K. U. Gross,
``Performance of one-body reduced density matrix functionals for the 
homogeneous electron gas'',
Phys. Rev. B \textbf{75}, 195120 (2007).

\bibitem{Callisto}
J. C. V\'arilly, J. M. Gracia-Bond\'{\i}a and W. Schempp,
``The Moyal representation of quantum mechanics and special function 
theory'',
Acta Appl. Math. \textbf{11}, 225--250 (1990).

\bibitem{Robert}
R. G. Littlejohn,
``The semiclassical evolution of wave packets'',
Phys. Rep. \textbf{138}, 193--291 (1986).

\bibitem{Titania}
J. M. Gracia-Bond\'{\i}a and J. C. V\'arilly,
``Nonnegative mixed states in Weyl--Wigner--Moyal theory'',
Phys. Lett. A \textbf{128}, 20--24 (1988).

\bibitem{BartlettM}
M. S. Bartlett and J. E. Moyal,
``The exact transition probabilities of quantum-mechanical
oscillators calculated by the phase-space method'',
Proc. Cambridge Phil. Soc. \textbf{45}, 545--553 (1949).

\bibitem{GritsenkoPB}
O. Gritsenko, K. Pernal and E. J. Baerends,
``An improved density matrix functional by physically motivated 
repulsive corrections'',
J. Chem. Phys. \textbf{122}, 204102 (2005).

\bibitem{Pluto}
Ph. Blanchard, J. M. Gracia-Bond\'{\i}a and J. C. V\'arilly,
``Density functional theory on phase space'',
physics.chem-ph/1011xxx. 

\bibitem{Rassolov}
V. A. Rassolov,
``An \textit{ab initio} linear electron correlation functional'',
J. Chem. Phys. \textbf{110}, 3672--3677 (1999).

\bibitem{GillCOB}
P. M. W. Gill, D. L. Crittenden, D. P. O'Neill and N. A. Besley,
``A family of intracules, a conjecture and the electron correlation 
problem'',
Phys. Chem. Chem. Phys. \textbf{8}, 15--25 (2006).

\end{thebibliography}
\end{document}